\documentstyle[12pt]{article}
\begin{document}

\tolerance=5000

\def\pp{{\, \mid \hskip -1.5mm =}}
\def\cL{{\cal L}}
\def\be{\begin{equation}}
\def\ee{\end{equation}}
\def\bea{\begin{eqnarray}}
\def\eea{\end{eqnarray}}
\def\beaa{\begin{eqnarray*}}
\def\eeaa{\end{eqnarray*}}
\def\tr{{\rm tr}\, }
\def\nn{\nonumber \\}
\def\e{{\rm e}}
\def\D{{D \hskip -3mm /\,}}

\  \hfill
\begin{minipage}{3.5cm}
OCHA-PP-171 \\
NDA-FP-90 \\
February 2001 \\
\end{minipage}

\vfill

\begin{center}
{\large\bf Quantum stabilization of thermal brane-worlds in M-theory}

\vfill

{\sc Shin'ichi NOJIRI}\footnote{nojiri@cc.nda.ac.jp},
{\sc Sergei D. ODINTSOV}$^{\spadesuit}$\footnote{
odintsov@ifug5.ugto.mx}, \\
and {\sc Sachiko OGUSHI}$^{\diamondsuit}$\footnote{
JSPS Research Fellow,
g9970503@edu.cc.ocha.ac.jp}
\\

\vfill

{\sl Department of Applied Physics \\
National Defence Academy,
Hashirimizu Yokosuka 239-8686, JAPAN}

\vfill

{\sl $\spadesuit$
Tomsk State Pedagogical University, 634041 Tomsk, RUSSIA and 
Instituto de Fisica de la Universidad de Guanajuato,
Lomas del Bosque 103, Apdo. Postal E-143, 
37150 Leon,Gto., MEXICO }

\vfill

{\sl $\diamondsuit$ Department of Physics,
Ochanomizu University \\
Otsuka, Bunkyou-ku Tokyo 112-0012, JAPAN}

\vfill

{\bf ABSTRACT}

\end{center}
 Low-energy effective M-theory (11d supergravity or 11d SUSY theory)
is considered in 11d flat or AdS background. After orbifold compactification
of eleventh dimension one gets flat 10d brane at non-zero temperature.
Quantum effective potential (free energy at low-temperature approximation)
 is calculated on such brane 
with account of only lowest mass from Kaluza-Klein modes. Such effective
 potential may stabilize the radius of 11th dimension (radion stabilization)
as it is demonstrated explicitly.

\newpage

Thermodynamics of extended objects is known to be quite remarkable.
It is marked by the presence of Hagedorn (critical) temperature 
for (super)strings\cite{strings} and (super)membranes\cite{membranes}.
It is expected that M-theory thermodynamics should also contain 
the Hagedorn temperature (for recent discussion, see \cite{Russo,ABP}).
From another side, brane-worlds \cite{RS} with trapping of gravity 
are expected to be useful in M-theory in different aspects \cite{HW}.
It could be that (classical and quantum) thermal effects in M-theory 
may provide the interesting consequences to brane-worlds which in their own
turn may help in better understanding of thermodynamics. In the recent 
model \cite{BMNO} it was demonstrated that thermal flat brane in 
AdS$_5$ bulk may be stabilized by quantum effects. In other words,
radion stabilization occurs due to thermal quantum effects.

In the present Letter we consider low-energy effective M-theory (11d 
supergravity or 11d SUSY theory) in flat or AdS 11d bulk. Thermal flat 
brane is obtained after orbifold $S_1/Z_2$ compactification of 
eleventh coordinate 
(the temperature is introduced in the standard way
 by compactification of time). The systematic method to calculate 
free energy (effective potential) at low-temperature approximation
 is developed with account of only lowest mass from KK tower\footnote{
The similar approach to calculation of graviton amplitudes in 11d SG 
via the corresponding evaluation of amplitudes in 10d SG with account of
 Kaluza-Klein modes has been developed in refs.\cite{GGV,RT}}.
The explicit calculation shows that such quantum free energy may stabilize 
the radius of orbifolded dimension. In other words, spontaneous 
compactification of 11d brane-worlds occurs due to account of thermal 
quantum effects.

We start with a simple review of black body 
radiation. Since the energy of a single particle with momentum 
${\bf p}$ and mass $m$ is given by
$E_p=\sqrt{{\bf p}^2 + m^2}$, 
 one obtains the following expression for its contribution to 
the partition function $Z_p^b$ for boson and $Z_p^f$ for 
fermion:
\be
\label{b23}
Z_p^b=\sum_{n=0}^\infty\e^{-\beta E_p\left( n+{1 \over 2}\right)}
={1 \over 2\sinh \left({\beta E_p \over 2}\right)}\ ,\ \ 
Z_p^f=\sum_{n=0}^1\e^{-\beta E_p\left( n - {1 \over 2}\right)}
=2\cosh \left({\beta E_p \over 2}\right)\ ,
\ee
 Here $\beta$ is the inverse temperature ,
$\beta= 1/T$, and we sum over all possible states with $n$ 
particles (quanta). In Eqs.(\ref{b23}) one  
expresses the zero-point energy by $\pm{1 \over 2}E_p$. 
The total partition function of the particle with mass $m$ 
in a $d-1$-dimensional volume $V_{d-1}$ is given 
by summing over $Z_k$ with respect to the momentum ${\bf p}$:
\be
\label{b45}
\beta F^{b, f}= -\ln Z^b = \pm V_{d-1}\int {d^{d-1}{\bf p} \over 
(2\pi)^{d-1}}\begin{array}{c}
\ln \left(2\sinh \left({\beta E_p \over 2}\right)\right) \\
\ln \left(2\cosh \left({\beta E_p \over 2}\right)\right)
\end{array}\ .
\ee
Here $F^{b,f}$ is free energy. 
 Sign $+$ and $\sinh$ ($-$ and $\cosh$) correspond to 
bosonic (fermionic) case in (\ref{b45}). 
The expr.~(\ref{b45}) diverge and require 
regularization in general. For 
supersymmetric case, we obtain a finite result:
\be
\label{b6}
\beta F^s = \beta\left(F^b + F^f\right)
= V_{d-1}\int {d^{d-1}{\bf p} \over 
(2\pi)^{d-1}}\ln \left(\tanh \left({\beta E_p \over 2}\right)
\right)\ .
\ee
The average energy $E$ is given by the derivative of the 
free energy, 
$E={\partial \over \partial \beta}\left(\beta F\right)$. 
Then for  above cases (\ref{b45}) and (\ref{b6}), 
one gets
\bea
\label{e2}
E^b_d(\beta;m)&=& V_{d-1}\int {d^{d-1}{\bf p} \over 
(2\pi)^{d-1}} {E_p \over 2} \coth \left(
{\beta E_p \over 2}\right)\ ,\\
\label{e3}
E^f_d(\beta;m)&=& -V_{d-1}\int {d^{d-1}{\bf p} \over 
(2\pi)^{d-1}}{E_p \over 2} \tanh \left({\beta E_p \over 2}\right)
\ ,\\
\label{e4}
E^s_d(\beta;m) &=& V_{d-1}\int {d^{d-1}{\bf p} \over 
(2\pi)^{d-1}} {E_p \over \sinh \beta E_p} \ .
\eea
This completes our elementary review of quantum statistics
at nonzero temperature.

As we will be mainly interested in the supersymmetric case, 
we subtract the zero temperature contributions
from $E^b_d(\beta;m)$ in Eq.~(\ref{e2}) and $E^f_d(\beta;m)$ 
in Eq.~(\ref{e3}) :
$\tilde E^{b,f}_d(\beta;m) = E^{b,f}_d(\beta;m)
 - E^{b,f}_d(\infty;m)$. 
(The zero temperature contributions in $E^b_d(\beta;m)$ 
cancel with each other in the supersymmetric case.)
 Note that $\tilde E^b_d(\beta;m)$ and 
$\tilde E^f_d(\beta;m)$ are finite. 
By changing the variable from $q$ to $s$ : 
$q=\sqrt{s^2 + 2\beta ms}$, 
the energies $\tilde E^b_d(\beta;m)$ and $\tilde E^f_d(\beta;m)$ 
have the following forms:
\bea
\label{l2}
\tilde E^{b,f}_d(\beta;m)={V_{d-1} \over 2^{d-1}\pi^{d-1 \over 2}
\Gamma\left({d-1 \over 2}\right)\beta^d}
\int_0^\infty ds {\left( s+\beta m \right)^2 
\left(s^2 + 2\beta ms\right)^{d-3 \over 2} 
\over \e^{s+\beta m} \mp 1} \ .
\eea
When $m=0$, one finds
\be
\label{m0}
\tilde E^b_d(\beta;0)={\pi^{d+1 \over 2}V_{d-1} \over 
\Gamma\left({d-1 \over 2}\right)\beta^d}
{B_{d \over 2} \over d}\ ,\quad 
\tilde E^f_d(\beta;0)={\pi^{d+1 \over 2}V_{d-1} \over 
\Gamma\left({d-1 \over 2}\right)\beta^d}
\left(1 - {1 \over 2^{{d \over 2}-1}}\right)
{B_{d \over 2} \over d} \ .
\ee
Here we assume $d$ is even and $B_n$ are the Bernoulli numbers. 
For example,  
$B_2={1 \over 30}$, $B_5={5 \over 66}$ 
and we used the following formulae
\be
\label{frml}
\int_0^\infty dx {2x^{2n-1} \over \e^{2\pi x} + 1}
={1 \over 2n}\left(1 - {1 \over 2^{n-1}}\right)B_n\ ,\quad
\int_0^\infty dx {2x^{2n-1} \over \e^{2\pi x} - 1}
={B_n \over 2n}\ .
\ee

Usually the temperature can be regarded to be much lower 
than the Planck scale. For example, even the characteristic 
temperature in hot inflationary universe is much lower. 
Then it will be enough if we work in the low temperature 
limit, where $\beta\rightarrow \infty$. Then one gets
\be
\label{l3}
\tilde E^{b,f}_d(\beta;m)\rightarrow
{V_{d-1} m^{d+1 \over 2}\e^{-\beta m} \over 
2^{d+1 \over 2}\pi^{d-1 \over 2}\Gamma\left({d-1 \over 2}\right)
\beta^{d-1 \over 2}}
\int_0^\infty ds s^{d-3 \over 2}\e^{-s} 
= {V_{d-1} m^{d+1 \over 2}\e^{-\beta m} \over 2
\left(2\pi\beta\right)^{d-1 \over 2}}\ .
\ee
The last expression has a maximum with respect to $m$ when 
$m={d+1 \over 2\beta}$. 
The expression (\ref{l3}) vanishes in the limit 
$m\rightarrow \infty$. The expression (\ref{l3}) also 
vanishes when $m=0$ but this is not exact since we take 
$\beta\rightarrow \infty$ limit first. The exact values 
when $m=0$ are given in Eqs.(\ref{m0}) and they are 
positive. 

We should sum up the energy with respect to the masses of 
the Kaluza-Klein modes, but due to 
the factor of $\e^{-\beta m}$ in (\ref{l3}), when $\beta$ 
is large it is enough to include only the lowest mass. 

 Let us consider M-theory, whose low energy 
effective theory is 11d supergravity, in 
the spacetime $R_{10}\times S_1$ or 
$R_{10}\times \left(S_1/Z_2\right)$ (see corresponding discussion in
\cite{HW}). 
The mass spectrum in the effective 10d theory 
is given by
\be
\label{Ms}
m^2={k^2 \over R^2} + l^2 R^2 \lambda^4\ .
\ee
Here $k$ and $l$ are integers, R is the radius of the 
 sphere $S_1$, and $\lambda^2$ is the tension. 
The second term corresponds to the mode where the membrane with 
the shape of a tube winds in $S_1$ if 11d theory is 
membrane. Since the situation does not 
 depend much on the spacetime dimension, one can consider  string 
 with one compactified dimension: $S_1$. 
The spectrum enjoys $T$-duality given by 
$R\rightarrow {1 \over \lambda^2 R}$. 
Therefore the lowest mass is 
\be
\label{lmass}
m=\left\{
\begin{array}{ll}
{1 \over R}\quad &  R\geq {1 \over \lambda} \\
\lambda^2 R \quad & R< {1 \over \lambda} \\
\end{array}\right.\ .
\ee
 Note that the lowest mass has maximum 
$m_{\rm max}=\lambda$ when $R={1 \over \lambda}$ 
when the energies (\ref{l3}) are:
\be
\label{l3b}
\tilde E^{b,f}_4(\beta;m)\sim
{V_{d-1} \lambda^{d+1 \over 2}\e^{-\beta \lambda} \over 2
\left(2\pi\beta\right)^{d-1 \over 2}}\ .
\ee
 One can identify the energy in the supersymmetric case 
as in (\ref{e4}) and effective potential with respect to the 
compactification radius $R$;
\be
\label{V}
V(R;\beta)=N\sum_{i=b,f}E_{10}^i\left(\beta; m(R)\right)
=N\sum_{i=b,f}\tilde E_{10}^i\left(\beta; m(R)\right)\ .
\ee
Here $N$ is number of bosonic degrees of freedom in supermultiplet. 
The potential has the local minima when $R=0$, $\infty$ ($m=0$) 
given by (\ref{m0}), or 
$R={1 \over \lambda}$ ($m=\lambda$) 
given by (\ref{l3b}). Due to the factor $\e^{-\beta \lambda}$, 
the minimum at $R={1 \over \lambda}$ is much lower than other 
two minima in the low energy temperature and becomes the global 
minimum. Therefore the thermal effects stabilize the radius of 
the compactification. 
Furtheremore by comparing (\ref{m0}) with (\ref{l3b}), 
we find the possibility of the phase transition when 
$\beta={1 \over \lambda}$. 
In the high temperature phase the radius vanishes or becomes 
infinite, which is equivalent due to $T$-duality. 
 Since we need to sum up all the Kaluza-Klein 
modes, which become massless in the limit of $R\rightarrow 0$ 
or $\infty$ and the quantum effects would be important when 
$\beta=\lambda$, we could not say any reliable things about the 
phase transition. 

Recently the free energy of the M-theory in the spacetime  
$R_{10}\times S_1$ or $R_9\times S_1\times S_1$ was 
calculated in \cite{Russo}. For $R_{10}\times S_1$, 
the obtained free energy is:
\bea
\label{Russo1}
{F(T) \over T}&=& - {945\zeta(11) \over 32\pi^5 R_{11}^{10}} 
 - {24\zeta(10) \over \pi^5 R_0^{10}} 
+ {\cal O}\left(\e^{-2\pi{R_0 \over R_{11}}}\right)\ ,
\quad R_{11}\ll R_0 \\
\label{Russo2}
{F(T) \over T}&=& - {945\zeta(11)R_{11} \over 32\pi^5 R_0^{11}} 
 - {24\zeta(10) \over \pi^5 R_0R_{11}^9} 
+ {\cal O}\left(\e^{-2\pi{R_{11} \over R_{10}}}\right)\ ,
\quad R_{11}\gg R_0 \ .
\eea
Here $R_{11}$ is the radius of $S_1$, $R_0=(2\pi T)^{-1}$ 
and $\zeta(x)$ is zeta function, especially 
$\zeta(10)={\pi^{10} \over 93555}$.
In \cite{Russo}, all the KK modes were summed up but the 
wrapping or winding mode has not been included. The 
 eqs.(\ref{Russo1}) and (\ref{Russo2}) were 
estimated for single boson. Therefore the first leading 
term in (\ref{Russo1}) should be cancelled in the 
supersymmetric theory.
 
In our
formulation, the low temperature 
approximation is applied. Only lowest 
massive modes are included in most of the regions, except the limit that 
all modes become massless, when $R\rightarrow 0$ or 
$R\rightarrow \infty$. In the limit $R\rightarrow 0$, the 
temperature dependent leading term of the supersymmetric free energy 
$F_{\rm R}$ in (\ref{Russo1} has the following form:
\be
\label{FRusso}
F_{\rm R}=-{24\cdot 2^{10} \pi^5 \zeta(10) V_9 \over \beta^{10}}
+ \begin{array}{l}\mbox{sub-leading} \\
\mbox{or $\beta$-independent terms}\end{array}\ .
\ee
Then the energy is given by
\be
\label{ERusso}
E_{\rm R}={\partial \over \partial\beta}\left(\beta F_{\rm R}
\right) 
={9\cdot 24\cdot 2^{10} \pi^5 \zeta(10) V_9 \over \beta^{10}}
+ \begin{array}{l}\mbox{sub-leading} \\
\mbox{or $\beta$-independent terms}\end{array}\ ,
\ee
which corresponds to the bosonic case in Eq.(\ref{m0}) with $d=10$. 
In Eq.(\ref{m0}), the 
contribution from only lowest massive mode was included but in 
(\ref{ERusso}), those from all massive modes were done. 
In spite of such a difference, the 
expression (\ref{ERusso}) is not so different from that in 
(\ref{m0}), that is, both of them are proportional 
to $\beta^{-10}$ and 
the coefficients are positive. Therefore the analysis given in 
this paper is valid even for $m\sim 0$.

As an extension, we now consider the orbifold compactification of 
 11d AdS space, as in the Randall-Sundrum model \cite{RS}. 
The 11th dimension, here called $y$, is compactified on an 
orbifold $S_1/Z_2$ of radius $R$, with  
$-\pi R\leq y \leq \pi R$. The orbifold fixed points, $y=0$ and 
$y=\pi R$, are the locations of two branes, and form the 
boundary of the 11-dimensional space-time. The 
11-dimensional metric is $g_{MN}$, with the 11d coordinates 
labelled by capital letters, $M=(\mu,11)$. The 10-dimensional 
Minkowski metric is $\eta_{\mu\nu}={\rm diag}(-1,1,1,\cdots,1)$, 
with $\mu=0$, $1$, $2$, $\cdots$, $9$. The line element is
\be
\label{Br1}
ds^2 = \e^{-2\sigma}\eta_{\mu\nu}dx^\mu dx^\nu + dy^2, 
\ee
where $\sigma=k |y|$. 
Here $k$ is a parameter of the Planck scale order, related 
to the AdS radius of curvature. The latter is equal to 
$1/k$. The points $(x^\mu,y)$ and $(x^\mu,-y)$ are 
identified. 
Let us assume that there is a real scalar field 
$\phi(x^\mu,y)$ in the bulk, with a mass $m_\phi$ given by
\be
\label{Br5}
m_\phi^2(y)=ak^2 + b\sigma''(y)\ ,
\ee
with $\sigma''(y)=2k \left[\delta(y) - \delta\sigma''(y)\right]$, 
$\sigma'(y)=k\epsilon (y) = k{y \over |y|}$. Here $a$ is an 
arbitrary non-dimensional parameter, defined as in \cite{GP}. 
The equation of motion for $\phi$ is
\be
\label{Br6}
{1 \over \sqrt{-g}} \partial_M\left(\sqrt{-g}g^{MN}
\partial_N\phi\right) - ak^2 \phi = 0\ ,
\ee
where $g=\det \left(g_{MN}\right)$. Using the metric 
(\ref{Br1}) one can write Eq.(\ref{Br6}) as 
\be
\label{Br7}
\left[\e^{2\sigma}\eta^{\mu\nu}\partial_\mu\partial_\nu 
+ \e^{10\sigma}
\partial_y\left(\e^{-10\sigma}\partial_y\right)
 - ak^2 \right]\phi(x^\mu,y)=0\ .
\ee
Expanding $\phi$ as
\be
\label{Br8}
\phi(x^\mu,y)={1 \over 2\pi R}\sum_{n=0}^\infty 
\phi^{(n)}(x^\mu)f_n(y)\ ,
\ee
we get field equations
\be
\label{Br10}
\eta^{\mu\nu}\partial_\mu \partial_\nu \phi^{(n)} 
= m_n^2 \phi^{(n)}\ , \quad 
 -{d \over dy}\left(\e^{-4\sigma} {df_n \over dy}\right) 
+ ak^2 \e^{-4\sigma} f_n = m_n^2 \e^{-2\sigma}f_n\ ,
\ee
in agreement with \cite{GP}. Here $f_n(y)$ are the 
Kaluza-Klein modes, and $m_n$ is the mass of the 
Kaluz-Klein excitation $\phi^{(n)}$. 
The solution of Eq.(\ref{Br10}) is \cite{GP}
\be
\label{Br12}
f_n(y)=\e^{5\sigma} \left[ J_\alpha\left(z_n\e^\sigma\right) 
+ b_\alpha\left(m_n\right) Y_\alpha\left(z_n\e^\sigma\right)
\right]\ ,
\ee
with $z_n={m_n \over k}$. Here $J_\alpha$, $Y_\alpha$ are 
the Bessel and Neumann functions, of order $\alpha=\sqrt{25+a}$. 
Assume that $\alpha\geq 0$. 
If the scalar field $\phi$ is even under $Z_2$ transformation: 
$y\rightarrow -y$, we have $f_n(y)=f_n(|y|)$. 
Then by using (\ref{Br5}), one finds 
$0=\left.\left({df_n \over dy} -b \sigma' f_n\right)
\right|_{y=0,\pi R}$ 
and we get the following two equations for 
determining $b_\alpha$ and $m_\alpha$:
\be
\label{Br156}
b_\alpha = {(5-b)J_\alpha(z_n) + z_n J_\alpha'(z_n) 
\over (5-b)Y_\alpha (z_n) + z_n Y_\alpha'(z_n) }\ , \quad 
b_\alpha(z_n)= b_\alpha \left(z_n\e^{\pi kR}\right)\ .
\ee
The above equations determine masses of the Kaluza-Klein modes. 
We only considered the masses of the scalar field $\phi$ 
but if the supersymmetry is preserved, the masses of the KK 
mode of spinor, which is the superpartner of $\phi$, should be 
identical with those of $\phi$. 

In order to investigate the qualitative structure of the roots 
in Eqs.(\ref{Br156}), one defines a new variable 
$x$ by $x\equiv z_n\e^{\pi kR}$ and rewrite Eqs. 
(\ref{Br156}) in the following 
form:
\be
\label{x1}
{(5-b)J_\alpha(x) + x J_\alpha'(x) 
\over (5-b)Y_\alpha (x) + x Y_\alpha'(x) }
={(5-b)J_\alpha\left(x\e^{-\pi kR}\right) + x\e^{-\pi kR} 
J_\alpha'\left(x\e^{-\pi kR}\right) 
\over (5-b)Y_\alpha \left(x\e^{-\pi kR}\right) + x\e^{-\pi kR} 
Y_\alpha'\left(x\e^{-\pi kR}\right) }\ .
\ee
 Note that there is a trivial solution $x=0$ in 
(\ref{x1}) but since the solution corresponds 
to the massless mode, we do not consider  it here. 
If we fix $x$ and take the limit of $R\rightarrow \infty$, 
the r.h.s. vanishes and in the limit
\be
\label{x2}
0=(5-b)J_\alpha(x) + x J_\alpha'(x)\ .
\ee
The lowest root $x$ of (\ref{x2}) is given by $x=0$, which 
corresponds to the massless mode when $R$ is finite and is 
neglected here.  Other nontrivial roots ($x\neq 0$) are finite 
and do not depend 
on $R$. Since the corresponding mass is given by 
\be
\label{x3}
m_n=kz_n=kx\e^{-\pi kR}\ ,
\ee
the non-trivial lowest mass vanishes in the limit, which 
corresponds to the local minimum in the potential 
(\ref{m0}). 

In order to investigate the behavior of Eq.(\ref{x1}) in 
the limit of $R\rightarrow 0$, one rewrites (\ref{x1}) in the 
following form:
\be
\label{x4}
0={{j_\alpha\left(x\e^{-\pi kR}\right) - j_\alpha (x) 
\over \e^{-\pi kR} - 1}y_\alpha(x) 
- j_\alpha(x){y_\alpha\left(x\e^{-\pi kR}\right) - y_\alpha (x) 
\over \e^{-\pi kR} - 1} 
\over y_\alpha(x)^2}\ .
\ee
Here
\be
\label{x5}
j_\alpha (x) \equiv (5-b)J_\alpha(x) + x J_\alpha'(x) \ ,\quad 
y_\alpha (x) \equiv (5-b)Y_\alpha(x) + x Y_\alpha'(x) \ .
\ee
Then in the limit of $R\rightarrow 0$, Eq.(\ref{x4}) becomes 
\be
\label{x6}
0={d \over dx}\left({j_\alpha(x) \over y_\alpha(x)}\right) \ .
\ee
Again the non-trivial lowest roots are finite and 
independent of $R$. Eq.(\ref{x3}) tells the mass is the order 
of the Planck mass since $x$ is of order unity. Then the mass 
corresponds to the global minimum of the 
effective potential at low temperature. 

The analysis in the limits $R\rightarrow \infty$ and 
$R\rightarrow 0$ suggests that the thermal effects would 
make the compactification radius to be small. 
If we include, however, the winding modes as in (\ref{Ms}), the 
corresponding masses vanish in the limit of $R\rightarrow 0$, 
which corresponds to the local minimum of the effective 
potential. This might suggest that there is a non-trivial maximum 
in the lowest mass as in flat spacetime in 
(\ref{lmass}). If $kR$ is equal or less than unity and 
$k$ has the order of the Planck mass, the maximum value of the 
lowest mass has the order of the Planck mass. 
Therefore the thermal effects might stabilize the 
compactification radius $R$.  In other words, the same thermal
 mechanism of radion 
stabilization which has been suggested in ref.\cite{BMNO} works 
(for other mechanisms, see \cite{mod}). It is quite natural as it is
fulfilled by quantum effects of fields from supermultiplet.  There is no 
  need to add extra fields.

 Let us consider non-supersymmetric case. If supersymmetry is 
broken, zero-temperature contributions from 
bosons and fermions are not cancelled with each other. Such zero 
temperature contributionse in $d=4$ thermal brane (AdS$_5$) have been 
evaluated in \cite{GR} (for related discussion of bulk Casimir effect 
in brane-worlds see refs.\cite{GPT, BMNO}).
 We now evaluate  vacuum energy for 
$d=10$ (AdS$_{11}$) case in a similar way. For zero 
temperature, (\ref{e2}) and (\ref{e3}) have 
the following forms:
\be
\label{zr1}
E_d^b(\infty; m)= - E_d^f(\infty; m) 
= {V_{d-1} \over 2}\int {d^{d-1} p \over (2\pi)^{d-1}}
\sqrt{p^2 + m^2 } 
= - {V_{d-1} \over 2(4\pi)^{d \over 2}}\Gamma
\left( -{d \over 2}\right)m^d\ .
\ee
 Suming up the contributions with respect to 
$m=kx\e^{-\pi k R}$ in (\ref{x3}) which 
satisfy (\ref{x1}) and using formula 
$A_s(\tilde a)\equiv \sum_n x_n^{-s}$ from  
\cite{GR}\footnote{It is convenient to use zeta-regularization in
such calculation, see \cite{EORBZ} for an introduction.} one gets:
\be
\label{zr2}
A_s(\tilde a)= {s \over \pi}\sin\left({\pi s \over 2}\right)
\int_0^\infty dt\,t^{-s-1} \ln \left[
{2 \e^{-t(1-\tilde a)} \over t\sqrt{\tilde a} }
\left\{k_\nu (t) i_\nu (\tilde a t) - k_\nu (\tilde a t) i_\nu (t)
\right\}\right]
\ee
Here $\tilde a=\e^{-\pi k R}$ and the Bessel functions of 
imaginary argument are
\be
\label{zr3}
i_\nu(z)=(5-b)I_\alpha(z) + zI_\alpha'(z) \ ,\quad 
k_\nu(z)=(5-b)K_\alpha(z) + zK_\alpha'(z)\ ,
\ee
which correspond to (\ref{x5}). If we choose 
$s=-d=-10 + \epsilon$ when $\epsilon\sim 0$, $A_s(a)$ has 
the following form: 
\bea
\label{zr4}
&& A_s(\tilde a)=5\epsilon\left\{\int_0^\infty dt\,t^{9-\epsilon}\ln 
\left[ 1 - {k_\alpha(t) i_\alpha (\tilde a t) \over  
k_\alpha(\tilde a t) i_\alpha (t)}\right] \right. \\
&& + \int_0^\infty dt\,t^{9-\epsilon}\ln 
\left[\sqrt{2\pi \over t}\e^{-t}i_\alpha(t)\right] 
\left. + {1 \over {\tilde a}^{10-\epsilon}} 
\int_0^\infty dt\,t^{9-\epsilon}\ln 
\left[-\sqrt{2 \over \pi t}\e^t k_\alpha(t)\right]\right\}\ .
\nonumber
\eea
As in \cite{GR}, the second term is cancelled by a local 
counterterm and the third term is absorbed into the 
renormalization of the Planck brane tension. 
Only the first term, which is finite, gives a non-trivial 
contribution to the effective potential. 
When $a$ is small
\bea
\label{zr5} 
\lefteqn{\int_0^\infty dt\,t^{9}\ln 
\left[ 1 - {k_\alpha(t) i_\alpha (\tilde a t) \over  
k_\alpha(\tilde a t) i_\alpha (t)}\right]} \nn
 \sim& {2 \over \alpha\Gamma(\alpha)^2}
{5-b + \alpha \over 5-b - \alpha}
\left({\tilde a \over 2}\right)^{2\alpha}
\int_0^\infty dt\,t^{9+2\alpha}{k_\alpha(t) 
\over i_\alpha(t)} \ &\mbox{for $5-b\neq \pm\alpha$} \\ 
\label{zr6} 
 \sim& {2 \over \alpha(\alpha + 1)\Gamma(\alpha)^2}
{7-b + \alpha \over 5-b - \alpha}
\left({\tilde a \over 2}\right)^{2\alpha+2}
\int_0^\infty dt\,t^{11+2\alpha}{k_\alpha(t) 
\over i_\alpha(t)} \ &\mbox{for $5-b=-\alpha$} \\ 
\label{zr7} 
 \sim& {2(-\alpha + 1) \over \alpha\Gamma(\alpha)^2}
{5-b + \alpha \over 7-b - \alpha}
\left({\tilde a \over 2}\right)^{2\alpha+2}
\int_0^\infty dt\,t^{11+2\alpha}{k_\alpha(t) 
\over i_\alpha(t)} \ &\mbox{for $5-b=\alpha$}\ .
\eea
As an example, we consider the case that $b=5$ and 
$\alpha> 0$. Then combining (\ref{zr1}), (\ref{zr4}) and 
(\ref{zr5}),  one gets bosonic (fermionic) 
 potential $V_0^b(\tilde a)$ and 
$V_0^f(\tilde a)$
\bea
\label{zr8} 
&& V_0^b(\tilde a)=-V_0^b(\tilde a)
=-V_9 k^{10}B_\alpha^{(0)} {\tilde a}^{10 + 2\alpha}\nn
&& B_\alpha^{(0)}\equiv -{2 \over 4! (2\pi)^5 
2^{2\alpha} \Gamma(\alpha)^2}
\int t^{9+2\alpha}{K_\alpha'(t) \over I_\alpha'(t)}>0\ .
\eea
 Including the leading low-temperature term 
 which corresponds to (\ref{l3}) one arrives to 
\be
\label{zr10}
V^{b,f}(\tilde a)=V_9\left[
\mp k^{10}B_\alpha^{(0)} {\tilde a}^{10 + 2\alpha} 
+ { (kxa)^5\e^{-\beta kxa} \over 2
\left(2\pi\beta\right)^{9 \over 2}}\right]\ .
\ee
Here sign $-$ ($+$) corresponds to boson (fermion). 
As we know that the $R$ or $x$ dependence  is small 
from the analysis in (\ref{x1}--\ref{x6}), we can regard $x$ as 
a constant of order unity. Then the second term in (\ref{zr10}) 
has a maximum of order $\beta^{-10}$ when $kax$ has order of 
$\beta^{-1}$. For $kax$ of order of $\beta^{-1}$, the first term in 
(\ref{zr10}) is about of $\beta^{-10}(\beta k)^{-2\alpha}$. Then 
for the fermionic case, the potential $V^f(\tilde a)$ 
(\ref{zr10}) has a nontrivial ($\tilde a>0$) minimum. 
If the value of the potential at the minimum is smaller 
than the value at $\tilde a=0$, which is given in the femionic 
case (\ref{m0}), 
the non-trivial minimum becomes a true minimum. Therefore the 
thermal effects can stabilize the compactification radius $R$. 
The value $a_{\rm min}$ of $a$ at the minimum would be same 
order as the value of $a$ at the maximum of the second 
term in (\ref{zr10}), then 
the order should be ${1 \over k\beta}$.  Taking the temperature 
as the order of the weak scale $\sim$ 10$^2$ GeV and assuming 
$k$ is about of the Planck scale $\sim$ 10$^{19}$ GeV, 
 one finds $a_{\rm min}\sim 10^{-17}$. 
This example proves that quantum bulk 
effects in a brane-world  AdS$_{11}$ at nonzero temperature may 
not only stabilize 10d brane-world (quantum spontaneous 
compactification occurs) but also provide
the dynamical mechanism \cite{BMNO} for the resolution of
 the hierarchy problem 
(with no fine-tuning) as in the Randall-Sundrum model \cite{RS}.
It is expected that similar effect occurs for complete effective 
field theory (11d SG) following from M-theory. It would be extremely
 interesting to extend this consideration for M(Membrane)-theory,
taking account of non-linear effects in order to get the corresponding 
Hagedorn temperature on curved 11d background.

\ 

\noindent
{\bf Acknoweledgements.} The work of S.D.O. was supported in 
part by CONACyT (CP, ref.9903056 and grant 28454E), by 
RFFI Grant N.99-02-16617 and by
GCFS Grant E00-3.3-461. The work of S.O. was supported 
in part by Japan Society for the Promotion of Science.

\end{document}